\documentclass[aps,twocolumn,amsmath,amssymb,superscriptaddress,preprintnumbers]{revtex4-2}
\usepackage[pdftex]{graphicx}
\pdfoutput=1
\usepackage[outdir=./]{epstopdf}
\usepackage{enumerate,ulem,color}

\newcommand{\modify}[1]{{#1}}

\newcommand{\beginsection}[1]{\noindent \textit{#1} --- }

\begin{document}
\title{Intrinsic torques emerging from anomalous velocity in magnetic textures}
\author{Yasufumi Araki}
\author{Jun’ichi Ieda}
\affiliation{Advanced Science Research Center, Japan Atomic Energy Agency, Tokai 319-1195, Japan}

\begin{abstract}
    \modify{
        Momentum-space topology of electrons under strong spin-orbit coupling contributes to the electrically induced torques exerting on magnetic textures
        insensitively to disorder or thermal fluctuation.
        We present a direct connection between band topology and the torques by classifying the whole torques phenomenologically.
        As well as the intrinsic anomalous Hall effect,
        the torques also emerge intrinsically from the anomalous velocity of electrons regardless of a nonequilibrium transport current.
    }
    We especially point out the intrinsic contribution arising 
    \modify{exclusively}
    in magnetic textures,
    which we call the ``topological Hall torque (THT)''.
    The THT emerges in bulk crystals without any interface or surface structures.
    We numerically demonstrate the enhancement of the THT in comparison with the conventional spin-transfer torque in the bulk metallic ferromagnet,
    \modify{which accounts for the giant current-induced torque measured in ferromagnetic $\mathrm{SrRuO_3}$.}
\end{abstract}

\maketitle

Nonuniform magnetic textures, such as domain walls (DWs) and skyrmions,
generate new functionalities in materials \cite{Soumyanarayanan_2016}.
\modify{Magnetic skyrmion} gives rise to the topological Hall effect (THE) \cite{Ye_1999,Tatara_2002,Lee_2009,Neubauer_2009,Kanazawa_2011},
\modify{which is described by the emergent electromagnetic fields associated with the geometric structure of electron wave functions in real space} \cite{Nagaosa_2012,Nagaosa_2013,Ieda_spincurrent,Tatara_2019}.
Aside from real space,
geometric structure in momentum space is also responsible for anomalous \modify{electron transport phenomena}.
The momentum-space Berry curvature leads to the anomalous velocity transverse to the applied electric field,
which is the origin of the intrinsic anomalous Hall effect (AHE) irrelevant to the nonequilibrium \modify{electron transport} \cite{Karplus_1954,Sundaram_1999,Nagaosa_2006,Sinistyn_2007,Nagaosa_2010}.
In the vicinity of band inversion point,
which typically occurs due to strong spin-orbit coupling (SOC)
on the surface of topological insulators (TIs) \cite{Nomura_2011,Yu_2010,Chang_2013,Checkelsky_2012} and in Weyl semimetals (WSMs) \cite{Grushin_2012,Goswami_2013,Burkov_2014,Burkov_2014_2,Nakatsuji_2015,Nakatsuji_2016,Nayak_2016,Liu_2018,Wang_2018},
the Berry curvature becomes significant and yields a large anomalous Hall conductivity (AHC).

\modify{Here
we explore cross effects spanning the two spaces focusing on} electrically induced torques that exert on magnetic textures under strong SOC.
The structure of torques was studied within particular models of TIs \cite{Garate_2010,Yokoyama_2010,Pesin_2012,Tserkovnyak_2012,Sakai_2014,Ndiaye_2017,Kurebayashi_2019,Imai_2021} and WSMs \cite{Kurebayashi_2016,Kurebayashi_2017,Kurebayashi_2019_2} in previous literature,
whereas universal understanding that is irrespective of their dispersions is not well established.
In the present study,
\modify{
    we point out that the anomalous velocity drives torques intrinsically as well as the AHE via spin-momentum locking (SML),
    by classifying the contributions phenomenologically. 
}
In comparison with the conventional spin-transfer torque (STT) \cite{Slonczewski_1989,Berger_1996,Tserkovnyak_2005} and spin-orbit torque (SOT) \cite{Obata_2008,Manchon_2008,Miron_2010,Miron_2011,Liu_2012,Liu_2012_2} induced by transport current,
the intrinsic torques are robust against disorder or thermal fluctuation.
Even in bulk crystals without breaking of inversion symmetry by surfaces or interfaces,
an intrinsic torque is generated by the coupling of momentum-space Berry curvature with real-space magnetic textures,
which we call the ``topological Hall torque (THT)''.
\modify{
    In ferromagnetic $\mathrm{SrRuO_3}$,
    we numerically demonstrate that the THT  scales around 10 times larger than the nonadiabatic STT,
    due to the SOC-induced Weyl points
    and the van Hove singularity (VHS) in the vicinity of them.
    It successfully accounts for the giant torque reported in the current-induced magnetization switching measurements \cite{Feigenson_2007,Yamanouchi_2019},
    whose mechanism was not identified in the previous literature.
}
\modify{
    Our theory of the intrinsic torques provides a universal guiding principle  
    to generate energy-efficient torques required for spintronics devices
    on the basis of band topology.
}

\beginsection{Classification of torques}
\modify{
    In order to point out topological contribution to the torques,
    we use the semiclassical (Boltzmann) formalism and classify the torques on the basis of symmetry.
    Here we consider two spin states of electrons
    under a strong SOC and a uniform magnetization with magnitude $M_s$ and direction $\boldsymbol{n}_0$.
    In the vicinity of the band inversion by SOC,
    the system is minimally
}
described by the two-band Hamiltonian
\begin{align}
    \hat{H}(\boldsymbol{k}) &= \boldsymbol{h}(\boldsymbol{k}) \cdot \hat{\boldsymbol{\sigma}}. \label{eq:two-band}
\end{align}
We denote operators as the characters with ``hat'' $(\hat{\mathcal{O}})$ throughout this article.
This model shows two eigenstates $\epsilon_\pm(\boldsymbol{k}) = \pm |\boldsymbol{h}(\boldsymbol{k})|$,
whose group velocities are given by $\boldsymbol{v}_\pm(\boldsymbol{k}) = \boldsymbol{\nabla}_{k} \epsilon_\pm(\boldsymbol{k})$. 
(We set $\hbar = 1$.)
This minimal model is capable of describing various band structures induced by SOC, 
such as Weyl nodes in WSMs and the Dirac spectrum on TI surfaces. 

Due to the strong SOC, the electron spin operator $\hat{\boldsymbol{\sigma}}$ is related to its velocity operator $\hat{\boldsymbol{V}}$ as
$    \hat{V}_i(\boldsymbol{k}) = \partial_{k_i} \hat{H}(\boldsymbol{k}) = \partial_{k_i} h_j \hat{\sigma}_j \equiv \Lambda_{ij}(\boldsymbol{k}) \hat{\sigma}_j,
$
where the tensor $\boldsymbol{\Lambda}$ characterizes the structure of SML. 
\modify{
    Therefore, the spin polarization and the velocity of an electron wave packet of momentum $\boldsymbol{k}$
    are related as $\boldsymbol{\sigma}(\boldsymbol{k}) = \boldsymbol{\Lambda}^{-1}(\boldsymbol{k}) \boldsymbol{V}(\boldsymbol{k})$,
    where $\boldsymbol{\sigma}(\boldsymbol{k})$ and $\boldsymbol{V}(\boldsymbol{k})$
    are the expectation values of $\hat{\boldsymbol{\sigma}}$ and $\hat{\boldsymbol{V}}$
    with respect to the wave function of a wave packet.
}

\modify{
    We now incorporate a real-space magnetic texture $\boldsymbol{n}(\boldsymbol{r}) = \boldsymbol{n}_0 + \delta\boldsymbol{n}(\boldsymbol{r})$,
    whose effect is described by the local additional term $\delta \hat{H}_{\mathrm{exc}}(\boldsymbol{r}) = \Gamma_{\mathrm{exc}} \delta \boldsymbol{n}(\boldsymbol{r}) \cdot \hat{\boldsymbol{\sigma}}$.
    Here we assume that the exchange coupling $\Gamma_{\mathrm{exc}}$ is isotropic.
} %
Then the effective magnetic field
\modify{
    that exerts a torque $\boldsymbol{t} = \gamma \boldsymbol{B}_{\mathrm{eff}} \times \boldsymbol{n}$ (with $\gamma$ the gyromagnetic ratio)
}
is given from the electron spin polarization as
\modify{
    $\boldsymbol{B}_{\mathrm{eff}}(\boldsymbol{r}) = -M_s^{-1} \langle \delta\hat{H}_{\mathrm{exc}}/\delta\boldsymbol{n} \rangle_{\boldsymbol{r}} = -(\Gamma_{\mathrm{exc}}/M_s) \langle \hat{\boldsymbol{\sigma}}\rangle_{\boldsymbol{r}}$,
    where $\langle\cdots\rangle_{\boldsymbol{r}}$ denotes the expectation value of the operator taken locally at position $\boldsymbol{r}$.
}
Therefore,
\modify{
    if the local electron distribution and the single-particle velocity for wave packets in each band $(\pm)$
    are given within semiclassical formalism as $F_\pm(\boldsymbol{r},\boldsymbol{k})$ and $\boldsymbol{V}_\pm(\boldsymbol{r},\boldsymbol{k})$, respectively,    
}
the effective field $\boldsymbol{B}_{\mathrm{eff}}(\boldsymbol{r})$ is given from them as
\begin{align}
    \boldsymbol{B}_{\mathrm{eff}}(\boldsymbol{r}) &= -\Delta \int [d\boldsymbol{k}] F(\boldsymbol{r},\boldsymbol{k}) \boldsymbol{\Lambda}^{-1}(\boldsymbol{k}) \boldsymbol{V}(\boldsymbol{r},\boldsymbol{k}), \label{eq:B-FLV}
\end{align}
with $\Delta = \Gamma_{\mathrm{exc}}/M_s$.
Here we omit the band indices $\pm$
and the summation over them,
and use the shorthand notation $[d\boldsymbol{k}] = d^d\boldsymbol{k}/(2\pi)^d$ with $d$ the spatial dimension.
\modify{
    Note that the modification of the phase space volume,
    due to the nontrivial topology in both real and momentum spaces \cite{Xiao_2006,Chang_2008,Xiao_2010},
    is incorporated in $F$ here for clarity of formulation.
}

\modify{
    Once we introduce an external electric field $\boldsymbol{E}$ and a magnetic texture $\boldsymbol{n}(\boldsymbol{r})$,
    the distribution $F$ and the velocity $\boldsymbol{V}$ deviate from those in the uniform and equilibrium system,
    $F_0 = f(\epsilon(\boldsymbol{k}))$ the Fermi distribution function and
    $\boldsymbol{V}_0 = \boldsymbol{v}(\boldsymbol{k})$.
}
In order to consider their effects on $\boldsymbol{B}_{\mathrm{eff}}$,
here we expand the quantities $Q=F,\boldsymbol{V},\boldsymbol{B}_{\mathrm{eff}}$ up to first orders in $\boldsymbol{E}$ and $\boldsymbol{\nabla}\boldsymbol{n}$,
which we denote symbolically as
\begin{align}
    Q \equiv Q_0 + \delta_E Q + \delta_{\nabla n} Q + \delta^2_{E\nabla n} Q. \label{eq:expansion}
\end{align}
\modify{
    $F$ is obtained from the Boltzmann equation,
    whereas $\boldsymbol{V}$ satisfies the semiclassical equations of motion for a wave packet \cite{Xiao_2006,Chang_2008,Xiao_2010}.
    Here we take the relaxation time $\tau$ for the Boltzmann equation,
    which phenomenologically accounts for scattering processes by disorder.
    $\boldsymbol{E}$ shifts $F$ by $\approx \tau e \boldsymbol{E} \cdot \boldsymbol{\nabla}_k F$,
    and hence $\delta_E F$ and $\delta^2_{E\nabla n} F$ are of $O(\tau^{\nu \geq 1})$.
    $\boldsymbol{E}$ also yields the anomalous velocity from momentum-space geometry,
    which corresponds to $\delta_E \boldsymbol{V}$ and $\delta^2_{E\nabla n} \boldsymbol{V}$.
}

With the expansion introduced above,
the contribution to the effective field $\boldsymbol{B}_{\mathrm{eff}}$ up to $O(\boldsymbol{E},\boldsymbol{\nabla} \boldsymbol{n})$ can be decomposed into 9 terms, which we summarize in Table \ref{table:classification}.
In particular, if the original system without $\boldsymbol{E}$ and $\boldsymbol{\nabla}\boldsymbol{n}$ satisfies inversion symmetry $\boldsymbol{h}(\boldsymbol{k}) = \boldsymbol{h}(-\boldsymbol{k})$,
we can determine the parity of each term in the integrand of Eq.~(\ref{eq:B-FLV})
under momentum inversion $\boldsymbol{k} \rightarrow -\boldsymbol{k}$,
which is displayed with the sign $\pm$ in each cell.
In such systems, the terms odd in $\boldsymbol{k}$ vanish under the momentum integral.

\begin{table}[tbp]
    \caption{Classification of the perturbative expansion of the effective field $\boldsymbol{B}_{\mathrm{eff}}$ in Eq.~(\ref{eq:B-FLV}).
    The expansion is given up to first orders in the electric field $\boldsymbol{E}$ and the magnetic texture $\boldsymbol{\nabla} \boldsymbol{n}$,
    as mentioned in Eq.~(\ref{eq:expansion}).
    The sign $\pm$ on each cell denotes the parity of each term in the integrand under momentum inversion $\boldsymbol{k} \rightarrow -\boldsymbol{k}$ in inversion-symmetric systems.
    The higher order terms beyond our present interest are displayed with $(\pm)$.
    Physical interpretations of the cells labeled with $a$-$d$ are given in the main text.}
    \label{table:classification}
    \begin{tabular}{|c|c||c|c|c|c|}
        \hline
        & & $F_0$ & $\delta_{\nabla n} F$ & $\delta_{E} F$ & $\delta^2_{E\nabla n} F$ \\ \hline
        $\boldsymbol{\Lambda}^{-1}$ & $-$ & $+$ & $-$ & $-$ & $+$ \\ \hline \hline
        $\boldsymbol{V}_0$ & $-$ & $+$ & $-$ & $^{b}\ -$ & $^{c}\ +$ \\ \hline
        $\delta_{\nabla n} \boldsymbol{V}$ & $+$ & $-$ & $(+)$ & $^{c}\ +$ & $(-)$ \\ \hline
        $\delta_{E} \boldsymbol{V}$ & $-$ & $^{a}\ -$ & $^{d}\ +$ & $(+)$ & $(-)$ \\ \hline
        $\delta^2_{E\nabla n} \boldsymbol{V}$ & $+$ & $^{d}\ +$ & $(-)$ & $(-)$ & $(+)$ \\ \hline
    \end{tabular}
\end{table}

Among the 9 terms in $\boldsymbol{B}_{\mathrm{eff}}$, the contribution at $O(\boldsymbol{E})$,
which is present even in the absence of magnetic texture,
is given by the two terms
\begin{align}
    \delta_{E}\boldsymbol{B}_{\mathrm{eff}} &= -\Delta \int [d\boldsymbol{k}] \left[ F_0 \ \boldsymbol{\Lambda}^{-1} \delta_{E} \boldsymbol{V} + \delta_{E} F \ \boldsymbol{\Lambda}^{-1} \boldsymbol{V}_0 \right],
\end{align}
which correspond to the cells $a$ and $b$ in Table \ref{table:classification}.
The first term, which we denote as $\delta_{E}^{(a)} \boldsymbol{B}_{\mathrm{eff}}$,
is the intrinsic contribution to the torque that is independent of electron transport.
It emerges from the anomalous velocity $\delta_{E} \boldsymbol{V}$,
which is induced by the momentum-space Berry curvature \cite{Karplus_1954,Sundaram_1999}.
The second term, which we denote as $\delta_{E}^{(b)} \boldsymbol{B}_{\mathrm{eff}}$,
gives the spin polarization in the electrically shifted distribution $\delta_{E} F$,
corresponding to the Rashba--Edelstein effect (REE) in systems under SOC \cite{Edelstein_1990,Belkov_2008}.
Both terms require breaking of inversion symmetry,
which is essential in the emergence of SOT.
If the system satisfies inversion symmetry,
the integrand becomes an odd function in $\boldsymbol{k}$,
and hence the above contributions vanish under the momentum integral.

Once inversion symmetry is broken by the magnetic texture,
the contribution to the torque becomes more diverse.
At $O(\boldsymbol{E},\boldsymbol{\nabla} \boldsymbol{n})$,
the contribution to $\boldsymbol{B}_{\mathrm{eff}}$ can be classified into two parts,
which correspond to the cells $c$ and $d$ in Table \ref{table:classification},
by considering whether the distribution is shifted by the electric field or not:
\begin{align}
    \delta^{2(c)}_{E\nabla n} \boldsymbol{B}_{\mathrm{eff}} &= -\Delta \!\! \int [d\boldsymbol{k}] \!\! \left[ \delta_{E} F \ \boldsymbol{\Lambda}^{-1} \delta_{\nabla n} \boldsymbol{V} + \delta^2_{E\nabla n} F \ \boldsymbol{\Lambda}^{-1} \boldsymbol{V}_0 \right] \nonumber \\
    \delta^{2(d)}_{E\nabla n} \boldsymbol{B}_{\mathrm{eff}} &= -\Delta \!\! \int [d\boldsymbol{k}] \!\! \left[ F_0 \ \boldsymbol{\Lambda}^{-1} \delta^2_{E\nabla n} \boldsymbol{V} + \delta_{\nabla n} F \ \boldsymbol{\Lambda}^{-1} \delta_{E}\boldsymbol{V} \right]. 
\end{align}
We can regard the former as the transport contribution,
since it comes from the nonequilibrium shift $\delta_E F$ or $\delta^2_{E\nabla n} F$,
\modify{
    which are of $O(\tau^{\nu \geq 1})$.
}
The latter can be identified as the geometric (or topological) contribution,
which emerges from the anomalous velocity $\delta_{E}\boldsymbol{V}$ or $\delta^2_{E\nabla n} \boldsymbol{V}$
and occurs at $O(\tau^0)$.
Since both $\boldsymbol{E}$ and $\boldsymbol{\nabla} \boldsymbol{n}$ are odd under spatial inversion,
the integrands become even functions in $\boldsymbol{k}$,
and hence these parts survive under the momentum integral even if the original system has inversion symmetry.

The geometric contribution $\delta^{2(d)}_{E\nabla n} \boldsymbol{B}_{\mathrm{eff}}$ originates from the anomalous velocity of an electron,
and also requires nonuniform magnetic textures.
Therefore,
we call this intrinsic torque the ``THT'',
in the sense similar to the AHE arising from the transverse anomalous velocity due to the momentum-space topology.
The THT is distinguished from the transport contribution,
including the conventional STT and SOT,
in that it is at $O(\tau^0)$ and hence is independent of a transport current.
While the transport contributions dominate in clean systems
where the perturbative expansion by $1/\tau$ works well,
the THT is also important because it is present even in dirty systems
with the short time scale of $\tau$, similarly to the AHE.


\beginsection{Torques in topological insulators}
Based on the conceptual classification of the torques shown above,
we demonstrate how the intrinsic torque emerges,
by taking an interface of a TI and a ferromagnet as a test case.
Since the TI surface shows SML 
in two dimensions,
we can apply the above formalism to in-plane spin textures,
but not to the out-of-plane component.
The aim of this discussion is not to establish an understanding on the detailed 
microscopic structure of torques therein,
which has been made intensively in several literatures \cite{Garate_2010,Yokoyama_2010,Tserkovnyak_2012,Sakai_2014,Ndiaye_2017,Kurebayashi_2019,Imai_2021},
but rather to demonstrate the emergence of the intrinsic torque phenomenologically.

We here take the out-of-plane magnetization as the ground state
and consider in-plane modulation of magnetic moments perturbatively.
The effective Hamiltonian for electrons at the interface is written as the massive 2D Dirac Hamiltonian at long wavelength,
$\hat{H}(\boldsymbol{k}) = c_F(k_x \hat{\sigma}_y - k_y \hat{\sigma}_x) + m \hat{\sigma}_z$,
where $m$ characterizes the exchange energy from the out-of-plane magnetization.
The Dirac spectrum becomes gapped and the electrons acquire the out-of-plane Berry curvature $\boldsymbol{\Omega}(\boldsymbol{k}) \parallel z$,
which leads to the quantized AHE \cite{Nomura_2011,Yu_2010,Chang_2013,Checkelsky_2012}.
This Hamiltonian shows the in-plane SML 
structure $\Lambda_{xy} = -\Lambda_{yx} = c_F$.

Due to the breaking of inversion symmetry at the interface,
the contributions to the torques at $O(\boldsymbol{E})$
exist even if the magnetization is uniform.
The electric field induces the anomalous velocity $\delta_E \boldsymbol{V} = e\boldsymbol{E}\times\boldsymbol{\Omega}$,
which yields the intrinsic torque independent of transport,
$\delta_E^{(a)} \boldsymbol{B}_{\mathrm{eff}} \parallel \boldsymbol{E}$.
This contribution
corresponds to an effect 
from the topological magnetoelectric coupling \cite{Qi_2008} identified previously \cite{Garate_2010,Tserkovnyak_2012,Ndiaye_2017}.
The electric field also shifts the distribution by $\delta_E F = \tau e \boldsymbol{E}\cdot\boldsymbol{v}f'(\epsilon)$,
which leads to the transport-related torque, $\delta_E^{(b)} \boldsymbol{B}_{\mathrm{eff}} \parallel \boldsymbol{e}_z \times \boldsymbol{E}$.
This contribution corresponds to the one from the conventional REE. 

In addition to the torques mentioned above,
we also have the contributions of $O(\boldsymbol{E},\boldsymbol{\nabla} \boldsymbol{n})$ in the presence of magnetic textures.
Since the exchange coupling to $\delta\boldsymbol{n}$ shifts the momentum $\boldsymbol{k}$ due to SML,
it can be effectively regarded as the U(1) gauge potential $\boldsymbol{a} = \Gamma_{\mathrm{exc}}\boldsymbol{\Lambda}^{-1}\delta\boldsymbol{n}/e$.
\modify{
    If $\boldsymbol{n}(\boldsymbol{r})$ has the N\'{e}el-type structure,
    $\boldsymbol{a}$ contains the magntic component, $b_z \equiv (\boldsymbol{\nabla}\times\boldsymbol{a})_z = -(\Gamma_{\mathrm{exc}}/e c_F)\boldsymbol{\nabla}\cdot\boldsymbol{n}$
    \cite{Nomura_2010,Tserkovnyak_2012,Wakatsuki_2015}.
    In this case, we should use the semiclassical formalism with a magnetic field
    to solve the dynamics of electrons \cite{Chang_1995,Chang_1996,Chang_2008,Xiao_2010}.
    This $b_z$
}
couples to the orbital magnetic moment $\mu_z(\boldsymbol{k}) = e\epsilon\Omega_z$ of a wave packet,
which leads to the shift $\delta_{\nabla n}\epsilon = -b_z \mu_z$ in the one-particle energy
and modifies the distribution by $\delta_{\nabla n}F$.
Due to such a breaking of inversion symmetry in $F$,
the anomalous velocity $\delta_E \boldsymbol{V}$ leads to the geometric contribution to the torque,
$\delta_{E\nabla n}^{2(d)}\boldsymbol{B}_{\mathrm{eff}} \propto (\boldsymbol{\nabla}\cdot\boldsymbol{n}) \boldsymbol{E}$,
which is the THT and behaves as the damping-like component
\modify{
    [see Sec.~S1.A of Supplemental Material (SM)]
}. 
This geometric contribution coexists 
with the transport contribution $\delta_{E\nabla n}^{2(c)}\boldsymbol{B}_{\mathrm{eff}}$.
It depends on the system parameters which contribution dominates:
since the 
microscopic analyses \cite{Sakai_2014,Kurebayashi_2019,Imai_2021} 
were based on the expansion by disorder scattering rate,
the transport contribution 
was dominantly treated and the geometric contribution including the THT is neglected in those analyses.
In contrast, the geometric part becomes dominant if the transport part is suppressed by disorder or thermal fluctuation.

\beginsection{\modify{THT in bulk ferromagnet with Weyl points}}
To highlight the THT as the dominant contribution to the torque, 
we consider a case that the contributions of $O(\boldsymbol{E})$ 
are suppressed by inversion symmetry.
Here we 
take the ferromagnetic system with Weyl fermions in the bulk as the test case and demonstrate the enhancement of the THT.
The effect of SOC and the momentum-space Berry curvature $\boldsymbol{\Omega}(\boldsymbol{k})$ become dominant around the Weyl points. 
By taking the model Hamiltonian satisfying inversion symmetry and SML simultaneously,
we numerically evaluate the THT.
We compare the THT with the AHE, which is known as the major consequence of  the momentum-space topology from the Weyl-node structure.
Since the AHC generally depends on the distance between two Weyl points \cite{Goswami_2013,Burkov_2014,Burkov_2014_2},
we here take a two-band model that can describe the pair structure of Weyl points \cite{Burkov_2011,Burkov_2011_2,Sharma_2016}.

\begin{figure}[bp]
    \includegraphics[width=8cm]{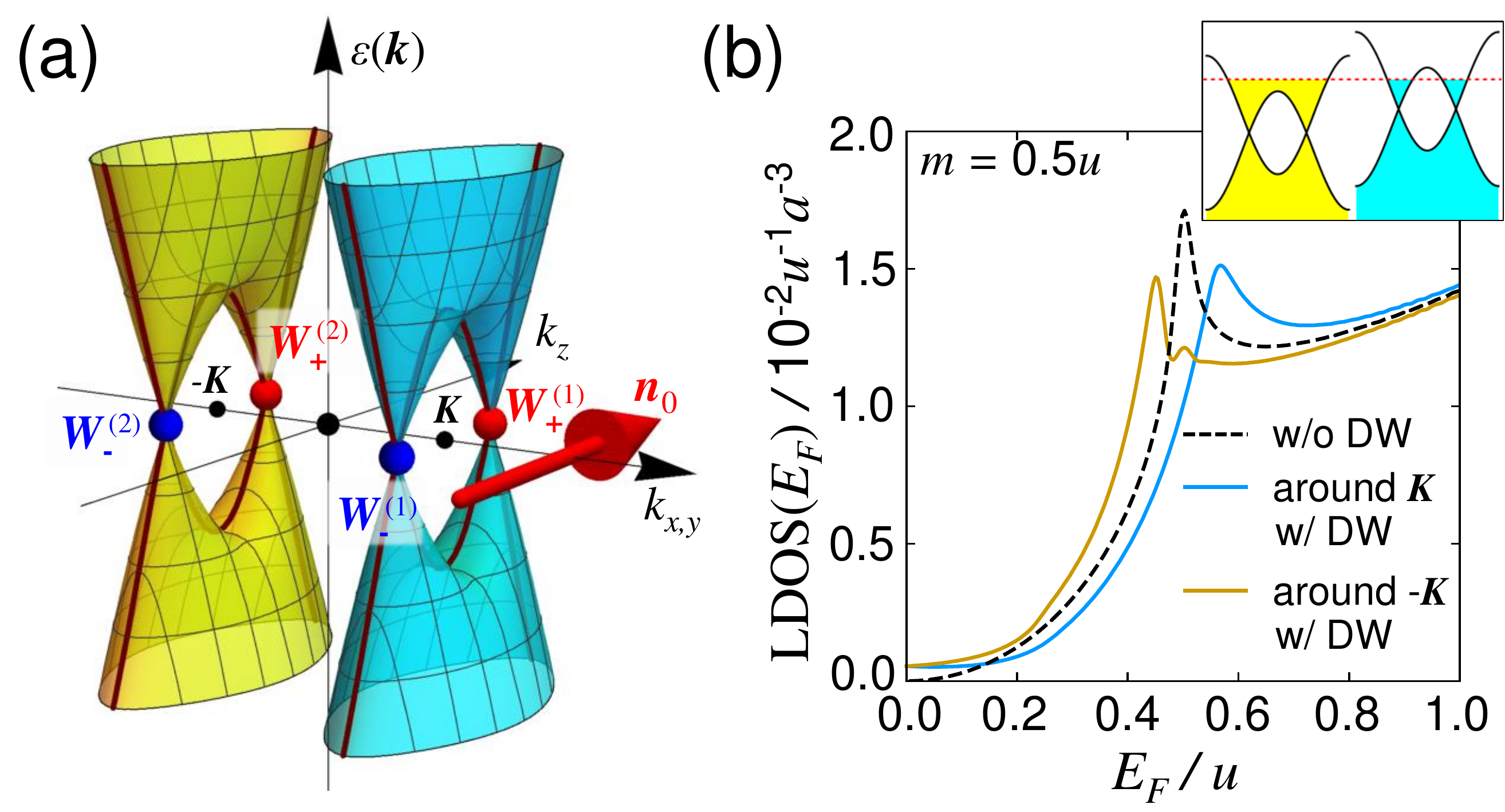}
    \caption{(a) Schematic picture of the band structure of the effective model with two pairs of Weyl points $(\boldsymbol{W}^{(1,2)}_\pm)$.
    (b) Local DOS calculated separately for the branches around $\boldsymbol{K}$ (cyan branch on panel (a)) and $-\boldsymbol{K}$ (yellow branch), from the effective model on lattice with $m=0.5u$.
    In the presence of a magnetic DW, the breaking of inversion symmetry leads to shift in the one-particle energy, as schematically depicted in the panel on the top, and the VHS in the DOS of each branch gets shifted oppositely to each other.}
    \label{fig:bands}
\end{figure}

If a pair of Weyl points are aligned along the $z$-axis around the momentum point $\boldsymbol{K}$,
the low-energy band structure around them is described by the two-band model with
$h_{x,y}(\boldsymbol{k}) = c_F (k_{x,y}-K_{x,y}), \quad
h_z(\boldsymbol{k}) = m-r(k_z-K_z)^2$,
where $r(>0)$ parametrizes the band inversion due to SOC. 
The parameter $m$ yields the splitting of Weyl points due to the breaking of time-reversal symmetry,
which usually corresponds to magnetization along the $z$-axis in magnetic Weyl (semi-)metals.
For $m>0$, the two Weyl points reside at momenta $\boldsymbol{W}_\pm^{(1)} = \boldsymbol{K} \pm \boldsymbol{e}_z\sqrt{m/r}$.
Since this single pair of Weyl points break spatial inversion symmetry,
we need another pair of Weyl points to preserve inversion symmetry,
to highlight the contribution at $O(\boldsymbol{E},\boldsymbol{\nabla} \boldsymbol{n})$ exclusively.
Inversion symmetry $\boldsymbol{h}(\boldsymbol{k}) = \boldsymbol{h}(-\boldsymbol{k})$ is restored by taking another pair of Weyl points at
$\boldsymbol{W}_\pm^{(2)} = - \boldsymbol{W}_\mp^{(1)} = -\boldsymbol{K} \pm \boldsymbol{e}_z\sqrt{m/r}$,
which is given from 
$h_{x,y}(\boldsymbol{k}) = -c_F (k_{x,y}+K_{x,y}), \quad
h_z(\boldsymbol{k}) = m-r(k_z+K_z)^2$ around $-\boldsymbol{K}$.
The band structure around the two pairs of Weyl points is schematically shown in Fig.~\ref{fig:bands}(a).
Note that the SML 
tensor $\boldsymbol{\Lambda}$ becomes diagonal around those Weyl points,
whereas it takes opposite signs for $\boldsymbol{K}$ and $-\boldsymbol{K}$
owing to inversion symmetry.
This model shows the VHS in the density of states 
(DOS) 
at momentum points $\pm\boldsymbol{K}$,
where the two Fermi 
pockets from the paired Weyl points 
merge into one.

We now consider the electrically induced torque on magnetic textures based on  the phenomenological classification in Table \ref{table:classification}.
For the magnetic texture $\delta \boldsymbol{n}(\boldsymbol{r})$ around $\boldsymbol{n}_0 \parallel z$,
we can apply the discussion similar to the case for TI heterostructure,
by introducing the effective gauge potential
$\boldsymbol{a} = \Gamma_{\mathrm{exc}} \boldsymbol{\Lambda}^{-1} \delta \boldsymbol{n} /e$
and the magnetic field $\boldsymbol{b} = \boldsymbol{\nabla} \times \boldsymbol{a}$.
Such a correspondence is the generalization of the ``axial gauge field'' picture
employed for magnetic Weyl semimetals with purely linear Weyl dispersion \cite{Kurebayashi_2017,Kurebayashi_2019_2,Liu_2013,Araki_2018,Araki_2020,Shitade_2021}.

The magnetic texture modifies the electron distribution $F$ via the magnetic component $\boldsymbol{b}$.
The shift in the one-particle energy $\delta_{\nabla n} \epsilon = -\boldsymbol{\mu} \cdot \boldsymbol{b}$ \cite{Chang_1995,Chang_1996,Xiao_2010} is odd under momentum inversion
as the consequence of the breaking of inversion symmetry by $\boldsymbol{\nabla} \boldsymbol{n}$.
Therefore, the distribution shift $\delta_{\nabla n} F$ also occurs oppositely around $\boldsymbol{K}$ and $-\boldsymbol{K}$.
The structure of $\delta_{\nabla n} F$ can be seen from the local 
DOS inside the DW,
which we calculate numerically on the hypothetical lattice model
\modify{
    (see Sec.~S2.B of SM).
}
In Fig.~\ref{fig:bands}(b), we show the local DOS from the bands around $\boldsymbol{K}$ and $-\boldsymbol{K}$ separately.
We can see that the energy of the VHS inside the DW gets shifted oppositely for the inversion partners,
as a consequence of the inversion symmetry breaking.
By incorporating also the modification of the integration measure by the factor $\tilde{D} = 1+e \boldsymbol{b} \cdot \boldsymbol{\Omega}$,
and the anomalous velocity $\delta_E \boldsymbol{V} = e\boldsymbol{E} \times \boldsymbol{\Omega}$ \cite{Chang_2008,Xiao_2010},
we obtain $\delta^{2(d)}_{E\nabla n} \boldsymbol{B}_{\mathrm{eff}}$ for the THT.

\begin{figure}[tbp]
    \includegraphics[width=6.5cm]{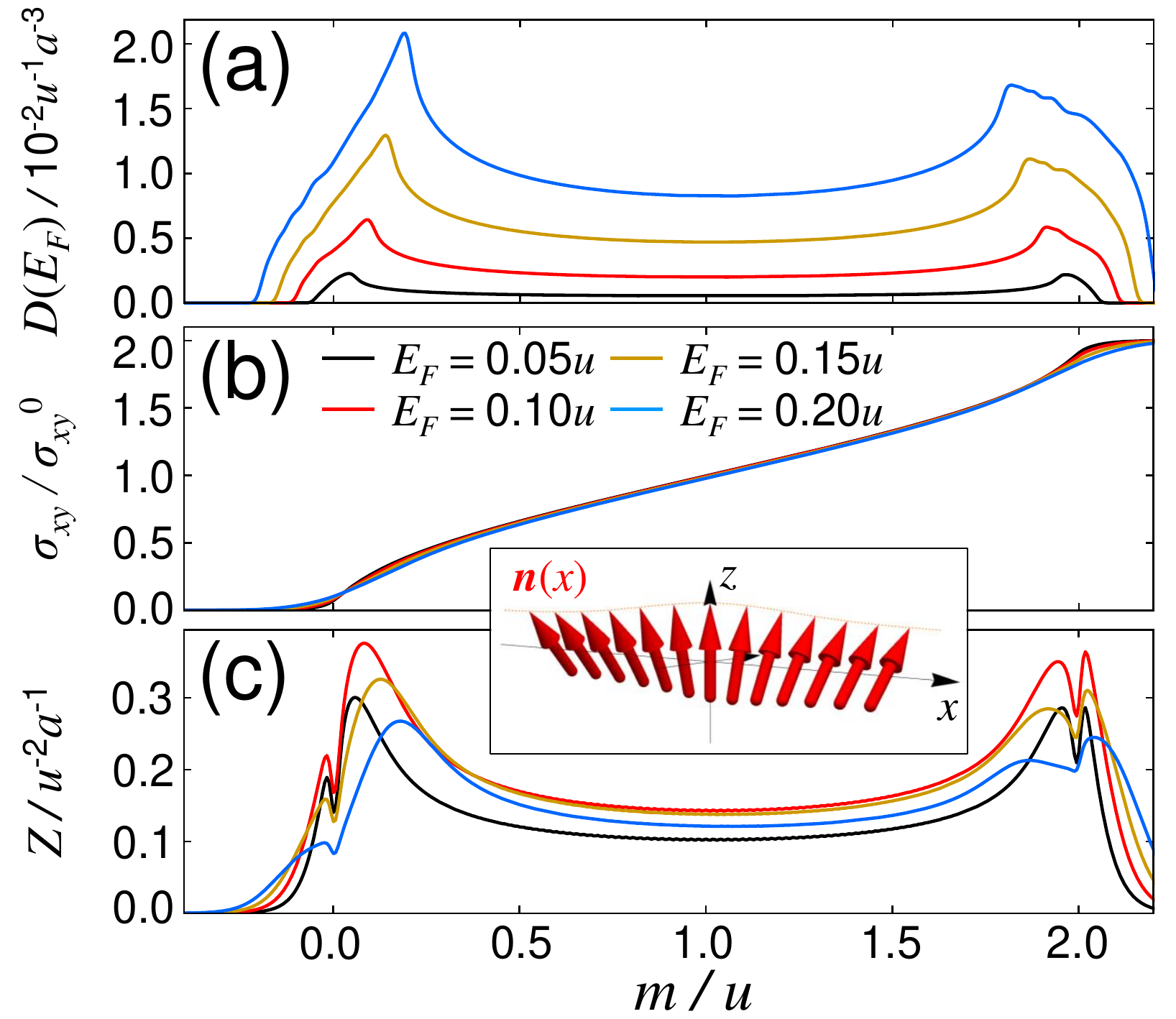}
    \caption{(a) The DOS $D(E_F)$, (b) the AHC $\sigma_{xy}$, and (c) the magnitude of the THT $Z$,
    calculated as a function of ``magnetization" $m$ with the lattice model of Weyl fermions.
    The calculations are carried out for the Fermi level $E_F = 0.05u$, $0.1u$, $0.15u$, and $0.2u$ with the 
    fixed temperature $T = 0.05u$. 
    $\sigma_{xy}^0 = e^2 / 2\pi a$ is the maximum value of the AHC that can be achieved by a single pair of Weyl points.
    The inset schematically shows the DW texture $\boldsymbol{n}(x)$ in this calculation.
    }
    \label{fig:calc}
\end{figure}

For concreteness,
if we take the Bloch-type magnetic texture $\boldsymbol{n}(x) = \boldsymbol{e}_z + \delta n_y(x) \boldsymbol{e}_y$
as shown in the inset of Fig.~\ref{fig:calc},
the effective field $\boldsymbol{b}$ points parallel to the $z$-axis,
and hence the THT induced by the electric field applied to $x$-direction $(E_x)$ reads
\begin{align}
    & \delta^{2(d)}_{E\nabla n} B^{\mathrm{eff}}_y = -\frac{\Gamma_{\mathrm{exc}}^2}{M_s} e E_x \partial_x n_y \int [d\boldsymbol{k}] \frac{f'(\epsilon)}{c_F^2}\epsilon \Omega_z^2. \label{eq:B-eff-Omega2}
\end{align}
\modify{
    (See Sec.~S1 in SM for detail of derivation.)
    Since the $\boldsymbol{k}$-space integral picks up the Berry curvature at the Fermi level,
    we generally need large Berry curvature and large DOS to realize a strong THT.
}
By denoting the factor $\int [d\boldsymbol{k}] \cdots $ as $-Z$,
which is determined only from the band structure,
the above form of the THT can be simplified as
$\delta^{2(d)}_{E\nabla n} \boldsymbol{B}_{\mathrm{eff}} = Z \frac{\Gamma_{\mathrm{exc}}^2}{M_s} e (\boldsymbol{E}\cdot \boldsymbol{\nabla}) \boldsymbol{n}$.
\modify{
    This form is similar to the nonadiabatic STT formulated in normal ferromagnetic metals,
    $\boldsymbol{B}_{\mathrm{eff}}^\beta = \frac{\beta P_s}{2 e \rho M_s}(\boldsymbol{E} \cdot \boldsymbol{\nabla}) \boldsymbol{n}$,
    where $P_s$ is the spin polarization rate,
    $\rho$ is the electric resistivity,
    and $\beta$ is the dimensionless parameter for nonadiabaticity \cite{Thiaville_2005}.
    The THT is compatible to $\boldsymbol{B}_{\mathrm{eff}}^\beta$ with $\beta = {2 Z \Gamma_{\mathrm{exc}}^2 e^2 \rho}/{P_s}(\equiv \beta_{\mathrm{THT}})$.
    The THT is allowed to exceed the transport-induced STT usually with $\beta \lesssim 0.1$ in normal metals,
    due to strong band topology and SOC.
}

\beginsection{Numerical 
evaluation of THT}
With the form of the THT obtained above,
we evaluate it numerically on a hypothetical cubic lattice with the lattice constant $a$.
Energy is measured by the unit $u = c_F a^{-1}$ on the lattice,
which roughly corresponds to the bandwidth (with some numerical factor),
and we fix the band inversion parameter $r=c_F a$.
By setting the temperature $T = 0.05u$ in the Fermi distribution function,
we numerically evaluate the DOS $D(E_F)$,
the intrinsic AHC $\sigma_{xy}$,
and the factor $Z$ for the THT,
for several values of the Fermi energy $E_F$ and the magnetization parameter $m$.
Their behaviors as functions of $m$ are shown in Fig.~\ref{fig:calc}.
As pointed out in previous analyses \cite{Grushin_2012,Goswami_2013,Burkov_2014,Burkov_2014_2},
$\sigma_{xy}$ depends monotonically on $m$ 
due to the Berry flux emerging between each pair of Weyl points.
In contrast, we see that the THT $(Z)$ gets largely enhanced at two points $m \approx 0,2u$.
It comes from the VHS in the DOS,
since the electron occupation around the VHS gets largely affected by the magnetic texture as we have seen above.
At the values of $m$ around the pair creation and annihilation of the Weyl points $(m=0,2u)$,
the VHS point in each Weyl point pair meets the Fermi level $E_F \approx 0$,
and hence the DOS and the THT are enhanced,
which is in clear contrast to the AHE.
Since the drift velocity $\boldsymbol{v} = \boldsymbol{\nabla}_k \epsilon$ vanishes at the VHS points due to the saddle-point structure in the bands,
the transport contribution to the torques tends to be suppressed around the VHS,
and the THT can make a dominant contribution.

\modify{
    Our model calculations can be applied to $\mathrm{Sr Ru O_3}$, which is a metallic ferromagnet with center symmetry.
    $\mathrm{Sr Ru O_3}$ is expected to have the Berry curvature and the SML structure around the Weyl points in its metallic bands \cite{Chen_2013},
    which consistently describe the large AHC \cite{Fang_2003} and spin-wave gap \cite{Itoh_2016} measured in its ferromagnetic phase.
    By using the material parameters of $\mathrm{Sr Ru O_3}$ listed up in Sec.~S2.C of SM (including Refs.~\cite{Nadgorny_2003,Raychaudhuri_2003,Rondinelli_2008}),
    the peak value $Z \approx 0.2 u^{-2}a^{-1}$ from the model calculation yields $\beta_{\mathrm{THT}} \approx 2$,
    which can hardly be reached by the transport-induced STT.
    The enhancement of current-induced DW motion measured in Refs.~\cite{Feigenson_2007,Yamanouchi_2019},
    whose origin has not been identified so far,
    can be successfully understood with this large $\beta_{\mathrm{THT}}$.
}

The THT contribution can exceed the conventional STT by ``engineering'' the band topology with magnetization \cite{Kurebayashi_2014,Ghimire_2019},
lattice strain \cite{strain}, etc.,
\modify{
    to pick up large Berry curvature and large DOS around the Fermi level.
}
We therefore emphasize the importance of bulk topology in spintronics,
in the sense that the proper tuning of bulk topology may make the device highly efficient
without building any complex heterostructures.


The authors thank Daichi Kurebayashi, Kentaro Nomura, and Michihiko Yamanouchi for fruitful discussions.
This work is partially supported by KAKENHI (no. 19H05622).
Y.A. is supported by the Leading Initiative for Excellent Young Researchers (LEADER).



\vspace{-12pt}
\begin{thebibliography}{99}
\vspace{-12pt}

\bibitem{Soumyanarayanan_2016}
A.~Soumyanarayanan,
N.~Reyren,
A.~Fert, and
C.~Panagopoulos,
\textit{Emergent phenomena induced by spin–orbit coupling at surfaces and interfaces},
Nature \textbf{539}, 509 (2016).

\bibitem{Ye_1999}
J.~Ye,
Y.~B.~Kim,
A.~J.~Millis,
B.~I.~Shraiman,
P.~Majumdar, and
Z.~Te\v{s}anovi\'{c},
\textit{Berry Phase Theory of the Anomalous Hall Effect: Application to Colossal Magnetoresistance Manganites},
Phys.~Rev.~Lett.~\textbf{83}, 3737 (1999).

\bibitem{Tatara_2002}
G.~Tatara and H.~Kawamura,
\textit{Chirality-Driven Anomalous Hall Effect in Weak Coupling Regime},
J.~Phys.~Soc.~Jpn. \textbf{71}, 2613 (2002).

\bibitem{Lee_2009}
M.~Lee,
W.~Kang,
Y.~Onose,
Y.~Tokura, and
N.~P.~Ong,
\textit{Unusual Hall Effect Anomaly in MnSi under Pressure},
Phys.~Rev.~Lett. \textbf{102}, 186601 (2009).

\bibitem{Neubauer_2009}
A.~Neubauer,
C.~Pfleiderer,
B.~Binz,
A.~Rosch,
R.~Ritz,
P.~G.~Niklowitz, and
P.~B\"{o}ni,
\textit{Topological Hall Effect in the A Phase of MnSi},
Phys.~Rev.~Lett. \textbf{102}, 186602 (2009).

\bibitem{Kanazawa_2011}
N.~Kanazawa,
Y.~Onose,
T.~Arima,
D.~Okuyama,
K.~Ohoyama,
S.~Wakimoto,
K.~Kakurai,
S.~Ishiwata, and
Y.~Tokura,
\textit{Large Topological Hall Effect in a Short-Period Helimagnet MnGe},
Phys.~Rev.~Lett. \textbf{106}, 156603 (2011).

%
%

%
%
%

%
%
%
%

\bibitem{Nagaosa_2012}
N.~Nagaosa and Y.~Tokura,
\textit{Emergent electromagnetism in solids},
Phys.~Scr.~\textbf{T146}, 014020 (2012).

\bibitem{Nagaosa_2013}
N.~Nagaosa and Y.~Tokura,
\textit{Topological properties and dynamics of magnetic skyrmions},
Nat.~Nanotechnol.~\textbf{8}, 899 (2013).

\bibitem{Ieda_spincurrent}
J.~Ieda and S.~Maekawa,
in \textit{Spin Current (Second Edition)}
(Eds: S.~Maekawa, S.~O.~Valenzuela, E.~Saitoh, and T.~Kimura,
Oxford University Press, Oxford, 2017, Ch.~7).

\bibitem{Tatara_2019}
G.~Tatara,
\textit{Effective gauge field theory of spintronics},
Physica E \textbf{106}, 208 (2019).

\bibitem{Karplus_1954}
R.~Karplus and
J.~M.~Luttinger,
\textit{Hall Effect in Ferromagnetics},
Phys.~Rev.~\textbf{95}, 1154 (1954).

\bibitem{Sundaram_1999}
G.~Sundaram and Q.~Niu,
\textit{Wave-packet dynamics in slowly perturbed crystals: Gradient corrections and Berry-phase effects},
Phys.~Rev.~B \textbf{59}, 14915 (1999).

\bibitem{Nagaosa_2006}
N.~Nagaosa,
\textit{Anomalous Hall Effect –A New Perspective–},
J.~Phys.~Soc.~Jpn.~\textbf{75}, 042001 (2006).

\bibitem{Sinistyn_2007}
N.~A.~Sinitsyn,
\textit{Semiclassical theories of the anomalous Hall effect},
J.~Phys.: Condens.~Matter \textbf{20}, 023201 (2007).

\bibitem{Nagaosa_2010}
N.~Nagaosa,
J.~Sinova,
S.~Onoda,
A.~H.~MacDonald, and
N.~P.~Ong,
\textit{Anomalous Hall effect},
Rev.~Mod.~Phys.~\textbf{82}, 1539 (2010).

\bibitem{Nomura_2011}
K.~Nomura and N.~Nagaosa,
\textit{Surface-quantized anomalous hall current and the magnetoelectric effect in magnetically disordered topological insulators},
Phys.~Rev.~Lett.~\textbf{106}, 166802 (2011).

\bibitem{Yu_2010}
R.~Yu,
W.~Zhang,
H.-J.~Zhang,
S.-C.~Zhang,
X.~Dai, and
Z.~Fang,
\textit{Quantized Anomalous Hall Effect in Magnetic Topological Insulators},
Science \textbf{329}, 61 (2010).

\bibitem{Chang_2013}
C.-Z.~Chang,
J.~Zhang,
X.~Feng,
J.~Shen,
Z.~Zhang,
M.~Guo,
K.~Li,
Y.~Ou,
P.~Wei,
L.-L.~Wang,
Z.-Q.~Ji,
Y.~Feng,
S.~Ji,
X.~Chen,
J.~Jia,
X.~Dai,
Z.~Fang,
S.-C.~Zhang,
K.~He,
Y.~Wang,
L.~Lu,
X.-C.~Ma, and
Q.-K.~Xue,
\textit{Experimental Observation of the Quantum Anomalous Hall Effect in a Magnetic Topological Insulator},
Science \textbf{340}, 167 (2013).

\bibitem{Checkelsky_2012}
J.~G.~Checkelsky,
J.~Ye,
Y.~Onose,
Y.~Iwasa, and
Y.~Tokura,
\textit{Dirac-fermion-mediated ferromagnetism in a topological insulator},
Nat.~Phys.~\textbf{8}, 729 (2012).

\bibitem{Grushin_2012}
A.~G.~Grushin,
\textit{Consequences of a condensed matter realization of Lorentz-violating QED in Weyl semi-metals},
Phys.~Rev.~D \textbf{86}, 045001 (2012).

\bibitem{Goswami_2013}
P.~Goswami and
S.~Tewari,
\textit{Axionic field theory of (3+1)-dimensional Weyl semimetals},
Phys.~Rev.~B \textbf{88}, 245107 (2013).

\bibitem{Burkov_2014}
A.~A.~Burkov,
\textit{Topological response in ferromagnets},
Phys.~Rev.~B \textbf{89}, 155104 (2014).

\bibitem{Burkov_2014_2}
A.~A.~Burkov,
\textit{Anomalous Hall Effect in Weyl Metals},
Phys.~Rev.~Lett.~\textbf{113}, 187202 (2014).

\bibitem{Nakatsuji_2015}
S.~Nakatsuji,
N.~Kiyohara, and
T.~Higo,
\textit{Large anomalous Hall effect in a non-collinear antiferromagnet at room temperature},
Nature \textbf{527}, 212 (2015).

\bibitem{Nakatsuji_2016}
N.~Kiyohara,
T.~Tomita, and
S.~Nakatsuji,
\textit{Giant Anomalous Hall Effect in the Chiral Antiferromagnet $Mn_3 Ge$},
Phys.~Rev.~Applied \textbf{5}, 064009 (2016).

\bibitem{Nayak_2016}
A.~K.~Nayak,
J.~E.~Fischer,
Y.~Sun,
B.~Yan,
J.~Karel,
A.~C.~Komarek,
C.~Shekhar,
N.~Kumar,
W.~Schnelle,
J.~K\"{u}bler,
C.~Felser, and
S.~S.~P.~Parkin,
\textit{Large anomalous Hall effect driven by a nonvanishing Berry curvature in the noncolinear antiferromagnet $Mn_3 Ge$},
Sci.~Adv.~\textbf{2}, e1501870 (2016).

\bibitem{Liu_2018}
E.~Liu,
Y.~Sun,
N.~Kumar,
L.~Muechler,
A.~Sun,
L.~Jiao,
S.-Y.~Yang,
D.~Liu,
A.~Liang,
Q.~Xu,
J.~Kroder,
V.~S\"{u}ss,
H.~Borrmann,
C.~Shekhar,
Z.~Wang,
C.~Xi,
W.~Wang,
W.~Schnelle,
S.~Wirth,
Y.~Chen,
S.~T.~B.~Goennenwein, and
C.~Felser,
\textit{Giant anomalous Hall effect in a ferromagnetic kagome-lattice semimetal},
Nat.~Phys.~\textbf{14}, 1125 (2018).

\bibitem{Wang_2018}
Q.~Wang,
Y.~Xu,
R.~Lou,
Z.~Liu,
M.~Li,
Y.~Huang,
D.~Shen,
H.~Weng,
S.~Wang, and
H.~Lei,
\textit{Large intrinsic anomalous Hall effect in half-metallic ferromagnet $Co_3 Sn_2 S_2$ with magnetic Weyl fermions},
Nat.~Commun.~\textbf{9}, 3681 (2018).

\bibitem{Garate_2010}
I.~Garate and M.~Franz,
\textit{Inverse Spin-Galvanic Effect in the Interface between a Topological Insulator and a Ferromagnet},
Phys.~Rev.~Lett.~\textbf{104}, 146802 (2010).

\bibitem{Yokoyama_2010}
T.~Yokoyama,
J.~Zang, and
N.~Nagaosa,
\textit{Theoretical study of the dynamics of magnetization on the topological surface},
Phys.~Rev.~B \textbf{81}, 241410(R) (2010).

\bibitem{Pesin_2012}
D.~Pesin and A.~H.~MacDonald,
\textit{Spintronics and pseudospintronics in graphene and topological insulators},
Nat.~Mater.~\textbf{11}, 409 (2012).

\bibitem{Tserkovnyak_2012}
Y.~Tserkovnyak and
D.~Loss,
\textit{Thin-Film Magnetization Dynamics on the Surface of a Topological Insulator},
Phys.~Rev.~Lett.~\textbf{108}, 187201 (2012).

\bibitem{Sakai_2014}
A.~Sakai and H.~Kohno,
\textit{Spin torques and charge transport on the surface of topological insulator},
Phys.~Rev.~B \textbf{89}, 165307 (2014).

\bibitem{Ndiaye_2017}
P.~B.~Ndiaye,
C.~A.~Akosa,
M.~H.~Fischer,
A.~Vaezi,
E.-A.~Kim,
and A.~Manchon
\textit{Dirac spin-orbit torques and charge pumping at the surface of topological insulators}
Phys.~Rev.~B \textbf{96}, 014408 (2017).

\bibitem{Kurebayashi_2019}
D.~Kurebayashi and N.~Nagaosa,
\textit{Theory of current-driven dynamics of spin textures on the surface of a topological insulator},
Phys.~Rev.~B \textbf{100}, 134407 (2019).

\bibitem{Imai_2021}
Y.~Imai,
T.~Yamaguchi,
A.~Yamakage, and
H.~Kohno,
\textit{Spintronic properties of topological surface Dirac electrons with hexagonal warping},
Phys.~Rev.~B \textbf{103}, 054402 (2021).

\bibitem{Kurebayashi_2016}
D.~Kurebayashi and K.~Nomura,
\textit{Voltage-Driven Magnetization Switching and Spin Pumping in Weyl Semimetals},
Phys.~Rev.~Applied \textbf{6}, 044013 (2016).

\bibitem{Kurebayashi_2017}
D.~Kurebayashi and K.~Nomura,
\textit{Microscopic theory of electrically induced spin torques in magnetic Weyl semimetals},
arXiv:1702.04918.

\bibitem{Kurebayashi_2019_2}
D.~Kurebayashi and K.~Nomura,
\textit{Theory for spin torque in Weyl semimetal with magnetic texture},
Sci.~Rep.~\textbf{9}, 5365 (2019).

\bibitem{Slonczewski_1989}
J.~C.~Slonczewski,
\textit{Conductance and exchange coupling of two ferromagnets separated by a tunneling barrier},
Phys.~Rev.~B \textbf{39}, 6995 (1989).

\bibitem{Berger_1996}
L.~Berger,
\textit{Emission of spin waves by a magnetic multilayer traversed by a current},
Phys.~Rev.~B \textbf{54}, 9353 (1996).

\bibitem{Tserkovnyak_2005}
Y.~Tserkovnyak,
A.~Brataas,
G.~E.~W.~Bauer,
and B.~I.~Halperin,
\textit{Nonlocal magnetization dynamics in ferromagnetic heterostructures},
Rev.~Mod.~Phys.~\textbf{77}, 1375 (2005).

\bibitem{Obata_2008}
K.~Obata and G.~Tatara,
\textit{Current-induced domain wall motion in Rashba spin-orbit system},
Phys.~Rev.~B \textbf{77}, 214429 (2008).

\bibitem{Manchon_2008}
A.~Manchon and S.~Zhang,
\textit{Theory of nonequilibrium intrinsic spin torque in a single nanomagnet},
Phys.~Rev.~B \textbf{78}, 212405 (2008).

\bibitem{Miron_2010}
I.~M.~Miron,
G.~Gaudin,
S.~Auffret,
B.~Rodmacq,
A.~Schuhl,
S.~Pizzini,
J.~Vogel, and
P.~Gambardella,
\textit{Current-driven spin torque induced by the Rashba effect in a ferromagnetic metal layer},
Nat.~Mater.~\textbf{9}, 230 (2010).

\bibitem{Miron_2011}
I.~M.~Miron,
K.~Garello,
G.~Gaudin,
P.-J.~Zermatten,
M.~V.~Costache,
S.~Auffret,
S.~Bandiera,
B.~Rodmacq,
A.~Schuhl, and
P.~Gambardella,
\textit{Perpendicular switching of a single ferromagnetic layer induced by in-plane current injection},
Nature \textbf{476}, 189 (2011).

\bibitem{Liu_2012}
L.~Liu,
O.~J.~Lee,
T.~J.~Gudmundsen,
D.~C.~Ralph,
and R.~A.~Buhrman,
\textit{Current-Induced Switching of Perpendicularly Magnetized Magnetic Layers Using Spin Torque from the Spin Hall Effect},
Phys.~Rev.~Lett.~\textbf{109}, 096602 (2012).

\bibitem{Liu_2012_2}
L.~Liu,
C.-F.~Pai,
Y.~Li,
H.~W.~Tseng,
D.~C.~Ralph, and
R.~A.~Buhrman,
\textit{Spin-Torque Switching with the Giant Spin Hall Effect of Tantalum},
Science \textbf{336}, 555 (2012).

\bibitem{Feigenson_2007}
M.~Feigenson,
J.~W.~Reiner, and
L.~Klein,
\textit{Efficient Current-Induced Domain-Wall Displacement in $SrRuO_3$},
Phys.~Rev.~Lett.~\textbf{98}, 247204 (2007).

\bibitem{Yamanouchi_2019}
M.~Yamanouchi,
T.~Oyamada,
K.~Sato,
H.~Ohta, and
J.~Ieda,
\textit{Current-Induced Modulation of Coercive Field in the Ferromagnetic Oxide $SrRuO_3$},
IEEE Trans.~Mag.~\textbf{55}, 1400604 (2019).

\bibitem{Xiao_2006}
D.~Xiao, Y.~Yao, Z.~Fang, and Q.~Niu,
\textit{Berry-Phase Effect in Anomalous Thermoelectric Transport},
Phys.~Rev.~Lett.~\textbf{97}, 026603 (2006).

\bibitem{Chang_2008}
M.-C.~Chang and Q.~Niu,
\textit{Berry curvature, orbital moment, and effective quantum theory of electrons in electromagnetic fields},
J.~Phys.: Condens.~Matter \textbf{20} 193202 (2008).

\bibitem{Xiao_2010}
D.~Xiao, M.-C.~Chang, and Q.~Niu,
\textit{Berry phase effects on electronic properties},
Rev.~Mod.~Phys.~\textbf{82}, 1959 (2010).

\bibitem{Gosselin_2006}
\modify{
    P.~Gosselin,
    F.~M\'{e}nas,
    A.~B\'{e}rard, and
    H.~Mohrbach,
    \textit{Semiclassical dynamics of electrons in magnetic Bloch bands: A Hamiltonian approach},
    Europhys.~Lett.~\textbf{76}, 651 (2006).
}

\bibitem{Edelstein_1990}
V.~M.~Edelstein,
\textit{Spin polarization of conduction electrons induced by electric current in two-dimensional asymmetric electron systems},
Solid State Commun.~\textbf{73}, 233 (1990).

\bibitem{Belkov_2008}
V.~V.~Bel'kov and
S.~D.~Ganichev,
\textit{Magneto-gyrotropic effects in semiconductor quantum wells},
Semicond.~Sci.~Technol.~\textbf{23}, 114003 (2008).

\bibitem{Qi_2008}
X.-L.~Qi,
T.~L.~Hughes, and
S.-C.~Zhang,
\textit{Topological field theory of time-reversal invariant insulators},
Phys.~Rev.~B \textbf{78}, 195424 (2008).

\bibitem{Nomura_2010}
K.~Nomura and
N.~Nagaosa,
\textit{Electric charging of magnetic textures on the surface of a topological insulator},
Phys.~Rev.~B \textbf{82}, 161401(R) (2010).

\bibitem{Wakatsuki_2015}
R.~Wakatsuki,
M.~Ezawa, and
N.~Nagaosa,
\textit{Domain wall of a ferromagnet on a three-dimensional topological insulator},
Sci.~Rep.~\textbf{5}, 13638 (2015).

\bibitem{Chang_1995}
M.-C.~Chang and Q.~Niu,
\textit{Berry Phase, Hyperorbits, and the Hofstadter Spectrum},
Phys.~Rev.~lett.~\textbf{75}, 1348 (1995).

\bibitem{Chang_1996}
M.-C.~Chang and Q.~Niu,
\textit{Berry phase, hyperorbits, and the Hofstadter spectrum: Semiclassical dynamics in magnetic Bloch bands},
Phys.~Rev.~B \textbf{53}, 7010 (1996).


\bibitem{Burkov_2011}
A.~A.~Burkov and
L.~Balents,
\textit{Weyl Semimetal in a Topological Insulator Multilayer},
Phys.~Rev.~Lett.~\textbf{107}, 127205 (2011).

\bibitem{Burkov_2011_2}
A.~A.~Burkov,
M.~D.~Hook, and
L.~Balents,
\textit{Topological nodal semimetals},
Phys.~Rev.~B \textbf{84}, 235126 (2011).

\bibitem{Sharma_2016}
G.~Sharma, P.~Goswami, and S.~Tewari,
\textit{Nernst and magnetothermal conductivity in a lattice model of Weyl fermions},
Phys.~Rev.~B \textbf{93}, 035116 (2016).

\bibitem{Liu_2013}
C.-X.~Liu,
P.~Ye, and
X.-L.~Qi,
\textit{Chiral gauge field and axial anomaly in a Weyl semimetal},
Phys.~Rev.~B \textbf{87}, 235306 (2013).

\bibitem{Araki_2018}
Y.~Araki and
K.~Nomura,
\textit{Charge pumping induced by magnetic texture dynamics in Weyl semimetals},
Phys.~Rev.~Applied \textbf{10}, 014007 (2018).

\bibitem{Araki_2020}
Y.~Araki,
\textit{Magnetic Textures and Dynamics in Magnetic Weyl Semimetals},
Ann.~Phys.~(Berlin) \textbf{532}, 1900287 (2020).

\bibitem{Shitade_2021}
A.~Shitade and
Y.~Araki,
\textit{Magnetization energy current in the axial magnetic effect},
Phys.~Rev.~B \textbf{103}, 155202 (2021).

\bibitem{Thiaville_2005}
\modify{
    A.~Thiaville,
    Y.~Nakatani,
    J.~Miltat, and
    Y.~Suzuki,
    \textit{Micromagnetic understanding of current-driven domain wall motion in patterned nanowires},
    Europhys.~Lett.~\textbf{69}, 990 (2005).
}

\bibitem{Chen_2013}
Y.~Chen, D.~L.~Bergman, and A.~A.~Burkov,
\textit{Weyl fermions and the anomalous Hall effect in metallic ferromagnets},
Phys.~Rev.~B \textbf{88}, 125110 (2013).

\bibitem{Fang_2003}
Z.~Fang,
N.~Nagaosa,
K.~S.~Takahashi,
A.~Asamitsu,
R.~Mathieu,
T.~Ogasawara,
H.~Yamada,
M.~Kawasaki,
Y.~Tokura,
and K.~Terakura,
\textit{The Anomalous Hall Effect and Magnetic Monopoles in Momentum Space},
Science \textbf{302}, 92 (2003).

\bibitem{Itoh_2016}
S.~Itoh, Y.~Endoh, T.~Yokoo, S.~Ibuka, J.-G.~Park, Y.~Kaneko, K.~S.~Takahashi, Y.~Tokura, and N.~Nagaosa,
\textit{Weyl fermions and spin dynamics of metallic ferromagnet SrRuO$_3$},
Nat.~Commun.~\textbf{7}, 11788 (2016).

\bibitem{Nadgorny_2003}
B.~Nadgorny,
M.~S.~Osofsky,
D.~J.~Singh,
G.~T.~Woods,
R.~J.~Soulen Jr.,
M.~K.~Lee,
S.~D.~Bu, and 
C.~B.~Eom,
\textit{Measurements of spin polarization of epitaxial SrRuO$_3$ thin films},
Appl.~Phys.~Lett.~\textbf{82}, 427 (2003).

\bibitem{Raychaudhuri_2003}
P.~Raychaudhuri,
A.~P.~Mackenzie,
J.~W.~Reiner, and
M.~R.~Beasley,
\textit{Transport spin polarization in SrRuO$_3$ measured through point-contact Andreev reflection},
Phys.~Rev.~B \textbf{67}, 020411(R) (2003).

\bibitem{Rondinelli_2008}
J.~M.~Rondinelli,
N.~M.~Caffrey,
S.~Sanvito, and
N.~A.~Spaldin,
\textit{Electronic properties of bulk and thin film SrRuO$_3$: Search for the metal-insulator transition},
Phys.~Rev.~B \textbf{78}, 155107 (2008).


\bibitem{Kurebayashi_2014}
D.~Kurebayashi and
K.~Nomura,
\textit{Weyl Semimetal Phase in Solid-Solution Narrow-Gap Semiconductors},
J.~Phys.~Soc.~Jpn.~\textbf{83}, 063709 (2014).

\bibitem{Ghimire_2019}
M.~P.~Ghimire,
J.~I.~Facio,
J.-S.~You,
L.~Ye,
J.~G.~Checkelsky,
S.~Fang,
E.~Kaxiras,
M.~Richter, and
J.~van den Brink,
\textit{Creating Weyl nodes and controlling their energy by magnetization rotation},
Phys.~Rev.~Research \textbf{1}, 032044(R) (2019)

\bibitem{strain}
See Refs.~\cite{Hirayama_2015,Ideue_2019,Shao_2017,Rancati_2020} for the topological phase transitions in nonmagnetic Dirac and Weyl semimetals.

\bibitem{Hirayama_2015}
M.~Hirayama,
R.~Okugawa,
S.~Ishibashi,
S.~Murakami, and
T.~Miyake
\textit{Weyl Node and Spin Texture in Trigonal Tellurium and Selenium},
Phys.~Rev.~Lett.~\textbf{114}, 206401 (2015).

\bibitem{Ideue_2019}
T.~Ideue,
M.~Hirayama,
H.~Taiko,
T.~Takahashi,
M.~Murase,
T.~Miyake,
S.~Murakami,
T.~Sasagawa, and
Y.~Iwasa,
\textit{Pressure-induced topological phase transition in noncentrosymmetric elemental tellurium},
Proc.~Natl.~Acad.~Sci.~USA \textbf{116}, 25530 (2019).

\bibitem{Shao_2017}
D.~Shao,
J.~Ruan,
J.~Wu,
T.~Chen,
Z.~Guo,
H.~Zhang,
J.~Sun,
L.~Sheng,
and D.~Xing,
\textit{Strain-induced quantum topological phase transitions in $Na_3Bi$},
Phys.~Rev.~B \textbf{96}, 075112 (2017).

\bibitem{Rancati_2020}
A.~Rancati,
N.~Pournaghavi,
M.~F.~Islam,
A.~Debernardi, and
C.~M.~Canali,
\textit{Impurity-induced topological phase transitions in $Cd_3As_2$ and $Na_3Bi$ Dirac semimetals},
Phys.~Rev.~B \textbf{102}, 195110 (2020).


\end{thebibliography}
\end{document}


\title{Supplemental Material for \\
``Intrinsic torques emerging from anomalous velocity in magnetic textures''}
\author{Yasufumi Araki}
\author{Jun’ichi Ieda}
\affiliation{Advanced Science Research Center, Japan Atomic Energy Agency, Tokai 319-1195, Japan}


\maketitle

\section{Evaluation of intrinsic torques under spin-momentum locking}
We here show the derivation process of the intrinsic torques under the concrete structure of spin-momentum locking $\boldsymbol{\Lambda}$ in the two-band model.
While we give our explanation in three dimensions (3D) for generality,
it is also applicable in 2D by limiting the dimensionality in $\boldsymbol{k}$ and $\boldsymbol{V}$.
After showing general discussion in 3D,
we demonstrate its application to the 2D interface of TI-ferromagnet and the 3D Weyl fermion system taken in the main text.

The starting point is the single-particle equations of motion
\begin{align}
    \dot{\boldsymbol{r}} &= \boldsymbol{\nabla}_k \tilde{\epsilon} - \dot{\boldsymbol{k}} \times \boldsymbol{\Omega} \label{eq:eom-r} \\
    \dot{\boldsymbol{k}} &= -e\boldsymbol{E} - e\dot{\boldsymbol{r}} \times \boldsymbol{b} \label{eq:eom-k}
\end{align}
for a wave packet \cite{Chang_1995,Chang_1996,Chang_2008,Xiao_2010}.
Here $\tilde{\epsilon}(\boldsymbol{k}) = \epsilon(\boldsymbol{k}) -\boldsymbol{\mu}(\boldsymbol{k})\cdot\boldsymbol{b}$ is the single-particle energy modified by the orbital magnetic moment $\boldsymbol{\mu}(\boldsymbol{k})$,
and $\boldsymbol{b}$ is the effective magnetic field corresponding to the magnetic texture $\boldsymbol{\nabla}\boldsymbol{n}$ introduced in the main text.
In order to obtain the perturbative expansion of the velocity $\boldsymbol{V}$,
we solve the equations of motion order by order in $\boldsymbol{E}$ and $\boldsymbol{b}$.
$\dot{\boldsymbol{k}}$ is given up to $O(\boldsymbol{E},\boldsymbol{b})$ as
\begin{align}
    \dot{\boldsymbol{k}} &= -e\boldsymbol{E} - e(\boldsymbol{\nabla}_k \epsilon - \dot{\boldsymbol{k}} \times \boldsymbol{\Omega}) \times \boldsymbol{b} \\
    &= -e\boldsymbol{E} - e(\boldsymbol{v} + e \boldsymbol{E} \times \boldsymbol{\Omega}) \times \boldsymbol{b}. \nonumber
\end{align}
By substituting this form to Eq.~(\ref{eq:eom-r}),
we obtain the perturbative expansion of the velocity $\boldsymbol{V}$,
\begin{align}
    \boldsymbol{V}_0 &= \boldsymbol{v} = \boldsymbol{\nabla}_k \epsilon(\boldsymbol{k}) \\
    \delta_E \boldsymbol{V} &= e \boldsymbol{E} \times \boldsymbol{\Omega} \\
    \delta_{\nabla n} \boldsymbol{V} &= -\boldsymbol{\nabla}_k (\boldsymbol{\mu}\cdot\boldsymbol{b}) +e(\boldsymbol{v}\times\boldsymbol{b})\times\boldsymbol{\Omega} \\
    \delta^2_{E\nabla n} \boldsymbol{V} &= e^2 [(\boldsymbol{E}\times\boldsymbol{\Omega})\times\boldsymbol{b}]\times\boldsymbol{\Omega} \\
    &= -e^2 (\boldsymbol{b}\cdot\boldsymbol{\Omega})(\boldsymbol{E}\times\boldsymbol{\Omega}). \nonumber
\end{align}

The distribution function $F$ in momentum space is given by
\begin{align}
    F(\boldsymbol{k}) &= \tilde{D}(\boldsymbol{k})g(\boldsymbol{k}),
\end{align}
which is composed of the single-particle distribution function $g(\boldsymbol{k})$,
and the correction factor $\tilde{D}(\boldsymbol{k}) = 1+e\boldsymbol{b}\cdot\boldsymbol{\Omega}$ for the integration measure (phase space volume) \cite{Xiao_2006,Chang_2008,Xiao_2010}.
The equilibrium distribution under the magnetic texture is described by the Fermi distribution function, $g(\boldsymbol{k}) = f(\tilde{\epsilon}(\boldsymbol{k}))$.
Therefore, the perturbative expansion of $F$ contributing to the geometric part is given as
\begin{align}
    F_0 &= f(\epsilon(\boldsymbol{k})) \\
    \delta_{\nabla n} F &= e (\boldsymbol{b}\cdot\boldsymbol{\Omega}) f(\epsilon) -(\boldsymbol{\mu}\cdot\boldsymbol{b})f'(\epsilon).
\end{align}
The perturbation in $F$ arising from the external electric field $\boldsymbol{E}$
tends to relax to the equilibrium distribution by the relaxation time $\tau$
and leads to the transport current,
and hence we identify the $\boldsymbol{E}$-dependent part as the transport contribution.

By substituting the perturbative expansions given above to the momentum integral
\begin{align}
    \boldsymbol{B}_{\mathrm{eff}} &= -\frac{\Gamma_{\mathrm{exc}}}{M_s} \int [d\boldsymbol{k}] F(\boldsymbol{k}) \boldsymbol{\Lambda}^{-1}(\boldsymbol{k}) \boldsymbol{V}(\boldsymbol{k}),
\end{align}
we obtain the intrinsic contributions to $\boldsymbol{B}_{\mathrm{eff}}$ as
\begin{align}
    \delta_{E}^{(a)} \boldsymbol{B}^{\mathrm{eff}} &= -\frac{\Gamma_{\mathrm{exc}}}{M_s} \int [d\boldsymbol{k}] f(\epsilon) \boldsymbol{\Lambda}^{-1} (e \boldsymbol{E} \times \boldsymbol{\Omega}) \\
    \delta^{2(d)}_{E\nabla n} \boldsymbol{B}^{\mathrm{eff}} &= \frac{\Gamma_{\mathrm{exc}}}{M_s} \int [d\boldsymbol{k}] f'(\epsilon)(\boldsymbol{\mu}\cdot\boldsymbol{b}) \boldsymbol{\Lambda}^{-1} (e \boldsymbol{E} \times \boldsymbol{\Omega}).
\end{align}
We use those two relations to evaluate the intrinsic torques for the TI heterostructure and the Weyl fermion system.

\subsection{TI-ferromagnet interface}
For the 2D interface of TI and ferromagnet,
we use the massive Dirac Hamiltonian $\hat{H}(\boldsymbol{k}) = c_F(k_x \hat{\sigma}_y - k_y \hat{\sigma}_x) + m \hat{\sigma}_z$,
with the spin-momentum locking structure $\Lambda_{xy} = -\Lambda_{yx} = c_F$.
This system shows the gapped spectrum $\epsilon(\boldsymbol{k}) = \sqrt{c_F^2|\boldsymbol{k}|^2 + m^2}$,
and the out-of-plane Berry curvature $\boldsymbol{\Omega}(\boldsymbol{k}) = -(c_F^2 m / 2\epsilon^3) \boldsymbol{e}_z$.
With those structures, the intrinsic torque at $O(\boldsymbol{E})$ is given as
\begin{align}
    \delta_{E}^{(a)} \boldsymbol{B}^{\mathrm{eff}} &= -\frac{\Gamma_{\mathrm{exc}}}{M_s} \int [d\boldsymbol{k}] f(\epsilon) \frac{\boldsymbol{e}_z}{c_F} \times (e \boldsymbol{E} \times \Omega_z \boldsymbol{e}_z) \\
    &= -\frac{\Gamma_{\mathrm{exc}}}{M_s} \int [d\boldsymbol{k}] f(\epsilon) \frac{e}{c_F} \Omega_z \boldsymbol{E}. \nonumber
\end{align}
At $O(\boldsymbol{E},\boldsymbol{\nabla}\boldsymbol{n})$, we use the effective magnetic field
\begin{align}
    \boldsymbol{b} &= \boldsymbol{\nabla} \times \boldsymbol{a} =  \boldsymbol{\nabla} \times \left(\Gamma_{\mathrm{exc}} \boldsymbol{\Lambda}^{-1} \delta \boldsymbol{n}/e \right) \\
    &= -\frac{\Gamma_{\mathrm{exc}}}{e c_F} (\boldsymbol{\nabla} \cdot\boldsymbol{n}) \boldsymbol{e}_z. 
    \quad\quad (\boldsymbol{\nabla} \cdot\boldsymbol{n} = \partial_x n_x + \partial_y n_y) \nonumber
\end{align}
\modify{
    The orbital magnetic moment defined for a wave packet \cite{Chang_1996,Chang_2008,Xiao_2010} is given by
    \begin{align}
        \boldsymbol{\mu}(\boldsymbol{k}) &\equiv -\frac{ie}{2} \langle \boldsymbol{\nabla}_k u(\boldsymbol{k}) | \times [\hat{H}(\boldsymbol{k}) - \epsilon(\boldsymbol{k})] | \boldsymbol{\nabla}_k u(\boldsymbol{k}) \rangle.
    \end{align}
    With the relations $\epsilon_+(\boldsymbol{k}) = -\epsilon_-(\boldsymbol{{k}})$ and $|u_+(\boldsymbol{k})\rangle \langle u_+(\boldsymbol{k}) | + |u_-(\boldsymbol{k})\rangle \langle u_-(\boldsymbol{k}) | =\boldsymbol{1}_{2 \times 2}$,
    the two-band Hamiltonian $\hat{H}(\boldsymbol{k})$ can be decomposed as
    \begin{align}
        \hat{H}(\boldsymbol{k}) &= \epsilon_+(\boldsymbol{k}) |u_+(\boldsymbol{k})\rangle \langle u_+(\boldsymbol{k}) | + \epsilon_-(\boldsymbol{k}) |u_-(\boldsymbol{k})\rangle \langle u_-(\boldsymbol{k}) | \nonumber \\
        &= \epsilon_+(\boldsymbol{k}) |u_+(\boldsymbol{k})\rangle \langle u_+(\boldsymbol{k})| \nonumber \\
        & \quad -\epsilon_+(\boldsymbol{k}) \left[\boldsymbol{1}_{2 \times 2}-|u_+(\boldsymbol{k})\rangle \langle u_+(\boldsymbol{k})|\right] \nonumber \\
        &= \epsilon_+(\boldsymbol{k}) \left[ 2|u_+(\boldsymbol{k})\rangle \langle u_+(\boldsymbol{k})| -\boldsymbol{1}_{2 \times 2} \right],
    \end{align}
    where we have restored the band indices $\pm$ for clarity of derivation.
    Therefore, $\boldsymbol{\mu}(\boldsymbol{k})$ reads
}
\begin{align}
    \boldsymbol{\mu}(\boldsymbol{k}) &= -\frac{ie}{2} \langle \boldsymbol{\nabla}_k u | \times 2\epsilon\left[|u\rangle \langle u| -\boldsymbol{1}_{2 \times 2} \right] | \boldsymbol{\nabla}_k u \rangle \nonumber \\
    &= ie \epsilon \langle \boldsymbol{\nabla}_k u | \times | \boldsymbol{\nabla}_k u \rangle - ie\epsilon \langle \boldsymbol{\nabla}_k u | u\rangle \times \langle u | \boldsymbol{\nabla}_k u \rangle \nonumber \\
    &= ie \epsilon \langle \boldsymbol{\nabla}_k u | \times | \boldsymbol{\nabla}_k u \rangle \nonumber \\
    &= e\epsilon(\boldsymbol{k}) \boldsymbol{\Omega}(\boldsymbol{k}).
\end{align}
for the two-band model.
By using those forms of $\boldsymbol{b}$ and $\boldsymbol{\mu}$, we obtain
\begin{align}
    \delta^{2(d)}_{E\nabla n} \boldsymbol{B}^{\mathrm{eff}} &= \frac{\Gamma_{\mathrm{exc}}^2}{M_s} (\boldsymbol{\nabla} \cdot\boldsymbol{n}) \int [d\boldsymbol{k}] \frac{e}{c_F^2} f'(\epsilon) \epsilon \Omega_z^2 \boldsymbol{E},
\end{align}
which is identified as the THT contribution.

\subsection{Weyl fermion system}
The model for the Weyl fermion system employed in this article has the linear spin-momentum locking structure
\begin{align}
    \Lambda_{xx} = \Lambda_{yy} = \pm c_F,
\end{align}
where the sign $\pm$ depends on the momentum regions around $\boldsymbol{K}$ or $-\boldsymbol{K}$ that the integration variable $\boldsymbol{k}$ belongs to.
If the energies of the bands in the intermediate region between $\boldsymbol{K}$ and $-\boldsymbol{K}$ are far from the band crossing region,
we can neglect the intermediate region and treat the contributions around $\boldsymbol{K}$ and $-\boldsymbol{K}$ separately.
Under this condition, we can treat the torques in a manner quite similar to the case for TI heterostructure.

By taking the Bloch-type magnetic texture $\boldsymbol{n}(x) = \boldsymbol{e}_z + \delta n_y(x) \boldsymbol{e}_y$,
the effective magnetic field becomes
\begin{align}
    \boldsymbol{b} &= \boldsymbol{\nabla} \times \left(\Gamma_{\mathrm{exc}} \boldsymbol{\Lambda}^{-1} \delta \boldsymbol{n}/e \right) =  \pm \frac{\Gamma_{\mathrm{exc}}}{e c_F} (\partial_x n_y) \boldsymbol{e}_z.  \nonumber
\end{align}
Note that the signs of $\boldsymbol{a}$ and $\boldsymbol{b}$ around $\boldsymbol{K}$ are opposite to those for $-\boldsymbol{K}$,
since $\boldsymbol{\Lambda}(\boldsymbol{k})$ is odd under momentum inversion due to inversion symmetry.
With the above form of $\boldsymbol{b}$, the THT contribution is given by
\begin{align}
    \delta^{2(d)}_{E\nabla n} \boldsymbol{B}^{\mathrm{eff}} &= -\frac{\Gamma_{\mathrm{exc}}}{M_s} \int [d\boldsymbol{k}]  \frac{\pm 1}{c_F} (e \boldsymbol{E} \times \boldsymbol{\Omega}) \\
    & \quad \quad \quad \quad \times f'(\epsilon) \left[- e \epsilon \boldsymbol{\Omega} \cdot \frac{\pm \Gamma_{\mathrm{exc}}}{e c_F} (\partial_x n_y) \boldsymbol{e}_z \right] \nonumber \\
    & = -\frac{\Gamma_{\mathrm{exc}}^2}{M_s} e E_x (\partial_x n_y) \boldsymbol{e}_y \int [d\boldsymbol{k}] \frac{f'(\epsilon)}{c_F^2}\epsilon \Omega_z^2,
\end{align}
as shown in the main text.

\section{Numerical calculation on lattice model}
We perform our numerical calculations with the Weyl fermions
by using the model defined on a hypothetical cubic lattice.
Here we first give the definition of our lattice model,
and then show our calculation method of the local DOS displayed in Fig.~1(b).

\subsection{Definition of lattice model}
We need to implement the low-energy effective Hamiltonian for two pairs of Weyl points,
\begin{align}
    h_x(\boldsymbol{k}) &= \pm c_F (k_x \mp K_x), \qquad
    h_y(\boldsymbol{k}) = \pm c_F (k_y \mp K_y), \nonumber\\
    h_z(\boldsymbol{k}) &= m - r(k_z \mp K_z)^2,
\end{align}
on the lattice,
where the sign $\pm$ labels the pair residing at $\boldsymbol{K}$ or $-\boldsymbol{K}$.
Since we are treating the contributions from each pair separately,
we may treat them with two independent lattice models.
Each pair of Weyl points can be reproduced by the lattice model
\begin{align}
    h_x(\boldsymbol{k}) &= \pm u \sin a k_x, \qquad
    h_y(\boldsymbol{k}) = \pm u \sin a k_y, \nonumber \\
    h_z(\boldsymbol{k}) &= m - u' \sum_{j = x,y,z} (1-\cos ak_j),
\end{align}
with the lattice parameters $u = c_F/a$ and $u' = 2r/a^2$.
Here we have shifted the Weyl point pair onto $k_z$-axis in each lattice model,
since the overall shift of the Weyl points does not matter in the numerical calculation.
We evaluate the momentum-space integrals for the anomalous Hall conductivity $\sigma_{xy}$ and the THT strength $Z$
on this lattice model,
with the mesh of $200 \times 200 \times 320$ points in the cubic Brillouin zone.
After evaluating the physical quantities with the two separate lattice models,
we add them up to obtain our final results.

\subsection{Calculation of local DOS}
The local DOS shown in Fig.~1(b) is calculated by setting up the DW structure on the lattice model shown above.
We add the exchange coupling term $\delta \hat{H}_{\mathrm{exc}} = \Gamma_{\mathrm{exc}} \delta \boldsymbol{n}(x) \cdot \hat{\boldsymbol{\sigma}}$ to the lattice Hamiltonian,
with the Bloch-type DW structure $\delta \boldsymbol{n}(x) = \Delta \tanh(x/w) \boldsymbol{e}_y$,
and evaluate the energy $\epsilon_{k_y,k_z,n}$ and wave function $\psi_{k_y,k_z,n}(x)$ for the eigenstates numerically.
(Note that the transverse momentum components $k_{y,z}$ serve as good quantum numbers under this DW structure.)
From the definition of the ordinary DOS
\begin{align}
    D(E_F) = \frac{1}{V} \sum_{k_y,k_z,n} \delta (\epsilon_{k_y,k_z,n} - E_F),
\end{align}
with $V$ the volume of the lattice system,
we define the local DOS at position $x$ by weighting the contribution from each state with its local amplitude $|\psi_{k_y,k_z,n}(x)|^2$,
\begin{align}
    D(E_F;x) = \frac{1}{V} \sum_{k_y,k_z,n} \delta (\epsilon_{k_y,k_z,n} - E_F) |\psi_{k_y,k_z,n}(x)|^2.
\end{align}
The calculation in Fig.~1(b) is carried out with the parameters
$\Gamma_{\mathrm{exc}} = u, \ \Delta = 0.25, \ w = 4a$,
and the local DOS with the DW is evaluated at the center of the DW $(x=0)$.

\subsection{Order estimation with $\mathrm{SrRuO_3}$}
The effective magnetic field from the nonadiabatic STT in normal ferromagnetic metal is given by
\begin{align}
    \boldsymbol{B}_{\mathrm{eff}}^\beta &= \frac{\beta P_s}{2 e M_s}(\boldsymbol{j} \cdot \boldsymbol{\nabla}) \boldsymbol{n},
\end{align}
where $P_s$ is the spin polarization rate,
$\boldsymbol{j}$ is the current density,
and $\beta$ is the dimensionless parameter for nonadiabaticity \cite{Thiaville_2005}.
Since the effective field for the THT is given by
\begin{align}
    \delta^{2(d)}_{E\nabla n} \boldsymbol{B}_{\mathrm{eff}} = Z \frac{\Gamma_{\mathrm{exc}}^2}{M_s} e (\boldsymbol{E}\cdot \boldsymbol{\nabla}) \boldsymbol{n}
\end{align}
from our analysis,
we can see that the THT is compatible with the nonadiabatic STT with
\begin{align}
    \beta_{\mathrm{THT}} = Z \frac{\Gamma_{\mathrm{exc}}^2 e\rho}{M_s} \frac{2 e M_s}{P_s}
    = \frac{2 Z \Gamma_{\mathrm{exc}}^2 e^2 \rho}{P_s},
\end{align}
where we have used the Ohm's law $\boldsymbol{E} = \rho \boldsymbol{j}$ with the longitudinal resistivity $\rho$.

The conversion of the THT strength $Z$ to the nonadiabaticity parameter $\beta_{\mathrm{THT}}$ is performed
by taking the following parameters of the metallic ferromagnet $\mathrm{Sr Ru O_3}$:
\begin{itemize}
    \item Longitudinal resistivity $\rho \approx 100 \; \mathrm{\mu\Omega\; cm}$ from the measurement at $T \approx 100 \; \mathrm{K}$ \cite{Yamanouchi_2019}.
    \item Spin polarization $P_s \approx 0.5$ from the measurements at $T \lesssim 4\;\mathrm{K}$ \cite{Nadgorny_2003,Raychaudhuri_2003}.
    \item Exchange energy $\Gamma_{\mathrm{exc}} \approx 0.6 \;\mathrm{eV}$ and the lattice constant $a \approx 4\;\text{\AA}$ given with the density-functional theory calculations \cite{Rondinelli_2008}.
    \item The Fermi velocity $c_F \approx 1\;\mathrm{eV}\text{\AA}$ typical in WSMs.
\end{itemize}
With those parameters, the THT strength $Z \approx 0.2$ obtained from our numerical calculation
corresponds to the nonadiabaticity parameter $\beta_{\mathrm{THT}} \approx 2$.
